\documentclass[aps,prd,twocolumn,preprintnumbers,floatfix,nofootinbib]{revtex4}
\usepackage{epsfig,epsf,graphics,psfrag,epstopdf}
\usepackage{times}
\usepackage{float}
\usepackage{color}
\usepackage{amsfonts,amssymb,stmaryrd,latexsym,amsmath}
\usepackage{multirow}
\usepackage{array} 
\usepackage{mathrsfs}
\usepackage{booktabs}
\usepackage{enumerate}
\usepackage{color}
\usepackage{subfigure}
\usepackage{hhline}
\usepackage[hyperindex=true]{hyperref}
\usepackage{epstopdf}
\usepackage{slashed}

\begin{document}
	
	\title{Radiative decays of $f_1(1285)$ as the $ K^*\bar K$ molecular state}

%
%
%
%
	
	\author{Ju-Jun Xie$^{1,4,5}$}
	\author{Gang Li$^{2,6}$}\email{gli@qfnu.edu.cn}
	\author{Xiao-Hai~Liu$^{3}$}

	\affiliation{
		$^1$Institute of Modern Physics, Chinese Academy of
		Sciences, Lanzhou 730000, China\\
		$^2$School of Physics and Engineering, Qufu Normal
		University, Shandong 273165, China\\
		$^3$Center for Joint Quantum Studies and Department of Physics, School of Science, Tianjin University, Tianjin 300350, China\\
		$^4$ School of Nuclear Science and Technology, University of Chinese Academy of Sciences, Beijing 100049, China\\
		$^5$School of Physics and Engineering, Zhengzhou University, Zhengzhou, Henan 450001, China\\
        $^6$ Institute of High Energy Physics and Theoretical Physics Center for Science Facilities, Chinese Academy of Sciences, Beijing 100049, China }

	\date{\today}
\begin{abstract}

Within a picture of the $f_1(1285)$ being a dynamically generated resonance from the $ K^*\bar K$ interactions, we estimate the rates for the radiative
transitions of the $f_1(1285)$ meson to the vector mesons $\rho^0$,
$\omega$ and $\phi$. These radiative decays proceed via the kaon loop diagrams. The calculated results are in fair agreement with the experimental measurements. Some predictions can be tested by experiments and their
implementation and comparison with these predictions will be
valuable to decode the nature of the $f_1(1285)$ state.

\end{abstract}

\maketitle
\raggedbottom

\section{Introduction}

The radiative decay mode of the $f_1(1285)$ resonance is interesting because it is the
basic element in the description of the $f_1(1285)$ photoproduction
data~\cite{Dickson:2016gwc,Osipov:2017ray}. It is also advocated as
one of the observables most suited to learn about the nature of the
$f_1(1285)$
state~\cite{Kalashnikova:2005zz,Nagahiro:2008cv,Oset:2016lyh,Geng:2015ifb,Dias:2017nwd,Tanabashi:2018oca}. By means of a chiral unitary approach, the $f_1(1285)$ appears as a pole in the complex plane of the scattering amplitude
of the $K^* \bar K+c.c.$ interaction in the isospin $I=0$ and $J^{PC} = 1^{++}$ channel~\cite{roca}. Or in another word, the axial-vector meson $f_1(1285)$ can be taken as a $K^*\bar{K}$ molecular state. For brevity, we use $K^* \bar K$ to represent the positive $C$-parity combination of $K^* \bar K$ and $\bar{K}^* K$ in the following parts. An
extension of the work of Ref.~\cite{roca}, including higher order
terms in the Lagrangian, has shown that the effect of the higher
order terms is negligible \cite{genghigher}. Using these theoretical
tools, predictions for lattice simulations in finite volume have
been done in Ref.~\cite{gengfinite}.


The experimental decay width of the $f_1(1285)$ is $22.7 \pm 1.1$
MeV~\cite{Tanabashi:2018oca}, quite small for its mass, and
naturally explained within the molecular state picture~\cite{roca}. The dominant decay
modes contributing to the width are peculiar. For example, the
$\eta \pi \pi$ channel accounts for $52\%$ of the width, and the
branching ratio of $\pi a_0(980)$ channel is $38\%$. The $\pi a_0(980)$ channel
has been well reproduced in Ref.~\cite{jorgifran} within the $K^*\bar{K}$
molecular state picture for the $f_1(1285)$, since the $a_0(980)$ strongly couples to $K\bar{K}$. In Ref.~\cite{jorgifran} the
$\pi f_0(980)$ decay mode was also studied, and the decay rate and
the invariant $\pi^+ \pi^-$ mass distribution were predicted. These
predictions have been confirmed in a recent BESIII
experiment~\cite{Ablikim:2015cob}. There is another important decay channel,
i.e. the $K \bar K \pi$, of which the branching ratio is $(9.1\pm 0.4)\%$~\cite{Tanabashi:2018oca}. This channel has ever been investigated in Ref.~\cite{Aceti:2015pma}
with the same picture as in Ref.~\cite{jorgifran}, and the theoretical calculations are compatible with the experimental measurements. As a matter of fact the success of $f_1(1285)$ as a  $K^* \bar K$ molecular state, being guided by the chiral unitary approach~\cite{roca}, has become more remarkable than before especially for its hadronic decay models. Yet, all the above test have been done in the hadronic decay modes and not in the radiative decays. This offers us the first opportunity to do this new test, which we conduct here.

On the experimental side, the Particle Data Group (PDG) averaged values on the radiative
decays of $f_1(1285)$ are~\cite{Tanabashi:2018oca}
\begin{eqnarray}
Br(f_1(1285) \to \gamma \rho^0) &=& (5.3 \pm 1.2) \%, \\
Br(f_1(1285) \to \gamma \phi) &=& (7.5 \pm 2.7) \times 10^{-4},
\end{eqnarray}
which lead to the partial decay width $\Gamma_{f_1(1285) \to \gamma
\rho^0} = 1.2 \pm 0.3$ MeV and a ratio $R_1 =
Br(f_1(1285) \to \gamma \rho^0)/Br(f_1(1285) \to \gamma \phi) = 71 \pm 30$.
There is currently no experimental data about the $f_1(1285) \to \gamma \omega$
decay. While the recent value of $\Gamma_{f_1(1285)
\to \gamma \rho^0}$ obtained by the CLAS Collaboration at Jafferson
Lab from the analysis of the $f_1(1285)$ photopruction off a proton
target is much smaller, which is $0.453 \pm
0.177$ MeV~\cite{Dickson:2016gwc}.
On the theoretical side, the authors in Ref.~\cite{Osipov:2017ray} give $\Gamma_{f_1(1285) \to \gamma \rho^0}=0.311$ MeV and $\Gamma_{f_1(1285) \to \gamma \omega}=0.0343$ MeV under the assumption that $f_1(1285)$ has a quark-antiquark nature. This $\Gamma_{f_1(1285) \to \gamma \rho^0}$ value is compatible with that of CLAS Collaboration within errors, but much smaller than the above PDG averaged value. Within the picture of $f_1(1285)$ being a quark-antiquark state, another theoretical prediction for the $f_1(1285)$ radiative decay is done in Ref.~\cite{Ishida:1988uw} using a covariant oscillator quark model. It predicts the $\Gamma_{f_1(1285) \to \gamma \rho^0}$ is in the range of $0.509\sim 0.565$ MeV, and $\Gamma_{f_1(1285) \to \gamma\omega}$ in the range of $0.024 \sim 0.057$ MeV, which depend on a particular mixing angle.


In this work, we extend the works of
Refs.~\cite{jorgifran,Aceti:2015pma} for the hadronic decays of
$f_1(1285)$ to the case of the radiative decays. In the molecular state scenario, the
$f_1(1285)$ decays into $\gamma V$ ($V =
\rho^0$, $\omega$, and $\phi$) via the kaon loop diagrams, and we can evaluate simultaneously these processes.  We show that the numerical results are in good agreement with
the experiment, hence supporting the molecular nature of the $f_1(1285)$ state.

The present paper is organized as follows: In sec.~\ref{sec:formalism}, we discuss the formalism and the main ingredients of the model; In sec.~\ref{sec:results} we present our numerical results and conclusions; A short summary is given in the last section.

\section{Formalism} \label{sec:formalism}

\begin{figure}[htbp]
\centering
\includegraphics[scale=0.48]{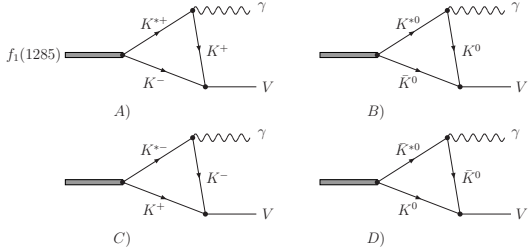}
\caption{Triangle loop diagrams representing the process
$f_1(1285) \to \gamma V$ with $V$ being the $\rho^0$, $\omega$, or $\phi$
meson.} \label{Fig:feydiagram}
\end{figure}

We study the decay of $f_1(1285) \to \gamma V$ with the assumption
that the $f_1(1285)$ is dynamically generated from the $K^* \bar{K}
+ c.c.$ interaction, thus this decay can proceed via $f_1(1285) \to
K^* \bar{K} \to \gamma V $ through the triangle loop
diagrams, which are shown in Fig.~\ref{Fig:feydiagram}. In this mechanism, the
$f_1(1285)$ first decays into $K^* \bar{K}$, then the $K^*$ decays into
$K \gamma$, and the $K\bar K$ interact to produce the vector
meson $V$ in the final state.
We use $p$, $k$, and $q$ for the momentum of $f_1(1285)$, $\gamma$ and
$K^-$ and $\bar{K}^0$ in Figs.~\ref{Fig:feydiagram} A) and B),
respectively. Then one can easily get the momentum of final vector meson is $p-k$,
and the momenta for $K^*$ and $K$ are $p-q$ and $p-q-k$, respectively.
\\
\\

In order to evaluate the partial decay width of $f_1(1285) \to
\gamma V$, we need the decay amplitudes of these diagrams shown in
Fig.~\ref{Fig:feydiagram}. As mentioned above, the $f_1(1285)$
resonance is dynamically generated from the interaction of
$K^* \bar{K}$. For the charge conjugate transformation, we take the phase conventions $\mathcal{C}K^* = -
\bar{K}^*$ and $\mathcal{C}K =
\bar{K}$, which are consistent with the standard chiral
Lagrangians, and write
\begin{eqnarray}
&& |f_1(1285)> = \frac{1}{\sqrt{2}} (K^* \bar{K} - \bar{K}^* K) \nonumber \\
& = & - \frac{1}{2} (K^{*+} K^- + K^{*0}\bar{K}^0
 - K^{*-}K^+ - \bar{K}^{*0} K^0)\ .
\end{eqnarray}
Then we can easily obtain the factors $C_1$ of $f_1\bar{K}K^*$
vertex for each diagram shown in Fig.~\ref{Fig:feydiagram},
\begin{eqnarray}
C^{A,B}_1 = - \frac{1}{2};\ \ C^{C,D}_1 = \frac{1}{2}.
\label{eq:factorc1}
\end{eqnarray}

For the $\bar K K V$ vertices, the effective Lagrangian describing the
vector-pseudoscalar-pseudoscalar ($VPP$)
interaction reads~\cite{hidden1,hidden2,hidden3,hidden4},
\begin{eqnarray}
{\cal L}_{VPP} = - i g <V^{\mu}[P,\partial_{\mu}P]>\ ,
\label{Eq:lvpp}
\end{eqnarray}
where $g = M/{2f} = 4.2$ with $M \approx (m_{\rho} + m_{\omega})/2$
and $f = 93$ MeV the pion decay constant. The
pseudoscalar- and vector-nonet are collected in the $P$ and $V$ matrices, respectively. The symbol $<>$ stands for
the trace.

According to the Lagrangian of Eq.~\eqref{Eq:lvpp}, the $\phi \to K\bar K$ decay width is given by
\begin{eqnarray}
\Gamma_{\phi \to K\bar K} = \frac{g^2m_{\phi}}{48\pi} \big( 1- \frac{4m^2_K}{m^2_{\phi}} \big )^{3/2}, \nonumber
\end{eqnarray}
and  we can obtain the coupling $g \simeq 4.5$ with the averaged experimental value of $\Gamma_{\phi \to K \bar K}$ in PDG~\cite{Tanabashi:2018oca}. We use $g=4.2$ in our calculations.

Thus, the
vertex of $\bar K K V$ can be written as
\begin{eqnarray}
\label{eq:vertex2} -i t_{\bar K K \to V} = i g C_2 (2q + k -
p)^{\mu} \varepsilon_{\mu}(p-k,\lambda_V),
\end{eqnarray}
where $\varepsilon_{\mu}(p-k,\lambda_V)$ is the polarization vector
of the vector meson. From Eq.~(\ref{Eq:lvpp}) and from the explicit
expressions of the $P$ and $V$ matrices, the factors $C_2$ for each
diagram shown in Fig.~\ref{Fig:feydiagram} can be obtained,
\begin{eqnarray}\label{C2-coeff}
C^{A,C}_2 &=& - \frac{1}{\sqrt{2}}; \ \ C^{B,D}_2 =
\frac{1}{\sqrt{2}}; ~~{\rm for ~ \rho ~production}, \nonumber\\
C^{A,C}_2 &=& - \frac{1}{\sqrt{2}}; \ \ C^{B,D}_2 =
-\frac{1}{\sqrt{2}}; ~~{\rm for ~ \omega ~production}, \nonumber\\
C^{A,C}_2 &=& 1; \ \ C^{B,D}_2 = 1; ~~{\rm for ~ \phi ~production}.
\end{eqnarray}
In terms of Eqs.~\eqref{eq:factorc1} and \eqref{C2-coeff}, it
is easy to know that Figs.~\ref{Fig:feydiagram} $A)$ and $C)$ give
the same contribution and Figs.~\ref{Fig:feydiagram} $B)$ and $D)$
also give the same contribution. We hence only consider
Figs.~\ref{Fig:feydiagram} $A)$ and $B)$ in the following
calculation.

For the electromagnetic vertex $K^* K \gamma$, the interaction takes the form~\cite{Oh:2006hm,Kim:2011rm,Kim:2012pz,Wang:2017tpe}
\begin{eqnarray}
{\cal L}_{K^*K\gamma} = \frac{eg_{K^*K\gamma}}{m_{K^*}}
\varepsilon^{\mu \nu \alpha \beta} \partial_\mu K^*_\nu
\partial_\alpha A_\beta K,
\end{eqnarray}
where $K^*_\nu$, $A_\beta$ and $K$ denote the $K^*$ vector meson,
photon, and the $K$ pseudoscalar meson, respectively. The partial decay width of $K^* \to K \gamma$ is given by
\begin{eqnarray}
\Gamma_{K^* \to K\gamma} = \frac{e^2g^2_{K^*K\gamma}}{96\pi}
\frac{(m^2_{K^*} - m^2_K)^3}{m^5_{K^*}}.
\end{eqnarray}
The values of the coupling constants $g_{K^*K\gamma}$ can be
determined from the experimental data~\cite{Tanabashi:2018oca},
which lead to
\begin{eqnarray}
g_{K^{*+}K^+ \gamma} = 0.75, ~~~~ g_{K^{*0}K^0 \gamma} = -1.14.
\end{eqnarray}
Here we fix the relative phase between the above two couplings taking into account the quark model expectation \cite{Close:1979bt}.

Here we give explicitly
the decay amplitude of Fig.~\ref{Fig:feydiagram} $A)$ for $\rho^0$
production,
\begin{eqnarray}\label{eq:MA}
M_{A} &=& - \frac{e g
g_{f_1}g_{K^{*+}K^+\gamma}}{2\sqrt{2}m_{K^{*+}}} \int
\frac{d^4q}{(2\pi)^4} \frac{1}{q^2 - m^2_{K^-} + i \epsilon} \nonumber \\
&\times & \frac{1}{2\omega^*(q)} \frac{D_1}{M_{f_1} -q^0 -
\omega^*(q) + i \Gamma_{K^{*+}}/2}  \nonumber \\
&\times & \frac{D_2}{(p-q-k)^2-m^2_{K^{+}} + i \epsilon} ,
\label{eq:ma}
\end{eqnarray}
where $\omega^*(q) = \sqrt{|\vec{q}~|^2 + m^2_{K^{*+}}}$ is the
$K^{*+}$ energy, and we have taken the positive energy part of the
$K^*$ propagator into account, which is a good approximation given
the large mass of the $K^*$ (see more details in
Ref.~\cite{jorgifran}). In Eq.~\eqref{eq:ma}, the factors $D_1$ and $D_2$ read
\begin{eqnarray}
D_1 &=& \varepsilon_{\mu \nu \alpha \beta} (p-q)^{\mu} \varepsilon^{\nu}(p,\lambda_{f_1}) k^{\alpha} \varepsilon^{*\beta}(k,\lambda_\gamma), \\
D_2 &=& (2q+k-p)^\sigma \varepsilon^*_{\sigma}(p-k,\lambda_\rho)\ ,
\end{eqnarray}
with $\lambda_{f_1}$, $\lambda_\gamma$, and $\lambda_{\rho}$ the
spin polarizations of $f_1(1285)$, photon and $\rho^0$ meson,
respectively. The amplitude $M_B$ corresponding to Fig.~\ref{Fig:feydiagram} $A)$ can be easily obtained through the substitutions $m_{K^{*+}} \to m_{K^{*0}}$, $m_{K^+} \to m_{K^0}$,
and $m_{K^-} \to m_{\bar{K}^0}$ in $M_A$. The decay amplitudes for $f_1(1285)\to\gamma\phi$ and $f_1(1285)\to\gamma\omega$ share the similar formalism as Eq.~\eqref{eq:MA}.

The partial decay width of the $f_1(1285) \to \gamma \rho^0$ decay
is given by
\begin{eqnarray}
\Gamma_{f_1(1285) \to \gamma \rho^0} = \frac{E_\gamma}{12\pi M^2_{f_1}}
\sum_{\lambda_{f_1}, \lambda_\gamma, \lambda_\rho} |M_A + M_B|^2.
\end{eqnarray}
The cases for $\omega$ and $\phi$ production
can be obtained straightforwardly.

To calculate $M_A$ in Eq.~\eqref{eq:ma}, we first integrate over $q^0$ using Cauchy's
theorem. For doing this, we take the rest frame of $f_1(1285)$, in
which one can write
\begin{eqnarray}
p &=& (M_{f_1},0,0,0), ~~ k=(E_\gamma,0,0,E_\gamma), \\
q &=& (q^0,|\vec{q}~|sin\theta cos\phi,|\vec{q}~|sin\theta
sin\phi,|\vec{q}~|cos\theta),
\end{eqnarray}
with $\theta$ and
$\phi$ the polar and azimuthal angles of $\vec{q}$ along the
$\vec{k}$ direction, and the energy of photon $E_\gamma = |\vec{k}~| = ({M^2_{f_1} -
m^2_{\rho^0}})/{2M_{f_1}}$. The energy of final vector meson is $E_V = M_{f_1} -E_\gamma$.
Then we have
\begin{eqnarray}
V_1 = D_1D_2 = \mp iE_\gamma |\vec{q}~|^2 sin^2\theta,
\end{eqnarray}
for $\lambda_{f_1} =0$, $\lambda_\gamma = \pm 1$, and
$\lambda_{\rho} = \mp1$, and
\begin{eqnarray}
V_2 &=& D_1D_2 = \pm i\frac{2E^2_\gamma}{m_{\rho^0}} \big ( q^0 -
M_{f_1} - |\vec{q}~|cos\theta \big ) \nonumber \\
&& \times (q^0 + \frac{E_V}{E_\gamma}
|\vec{q}~|cos\theta),
\end{eqnarray}
for $\lambda_{f_1} =\pm 1$, $\lambda_\gamma = \pm 1$, and
$\lambda_{\rho} = 0$. Notice that we have dropped those terms
containing $sin\phi$ or $cos\phi$, because after the integration over
$\phi$, they do not give contributions.

After integrating over $q^0$ in Eq.~\eqref{eq:ma}, we
have
\begin{widetext}
\begin{eqnarray}
F^A_1 &=& \frac{|\vec{q}~|^4 (1-cos^2\theta)}{\omega\omega'\omega^*}
\big ( X^A_1 + X^A_2 + X^A_3 \big ), \\
F^A_2 &=& \frac{|\vec{q}~|^2}{\omega\omega'\omega^*} \big [(M_{f_1}
-\omega^* + \frac{E_V}{E_\gamma}|\vec{q}~|cos\theta) (-\omega^* -
|\vec{q}~|cos\theta) X^A_1 + (\omega - M_{f_1} -
|\vec{q}~|cos\theta)  \nonumber \\
&& \times  (\omega + \frac{E_V}{E_\gamma}|\vec{q}~|cos\theta)X^A_2 + (\omega' - E_\gamma
- |\vec{q}~|cos\theta) (E_V + \omega' + \frac{E_V}{E_\gamma}|\vec{q}~|cos\theta) X^A_3 \big ],
\end{eqnarray}
\end{widetext}
where
\begin{eqnarray}
X^A_1 &=& \frac{1}{(M_{f_1} - \omega^* - \omega +
i\epsilon)(E_\gamma
- \omega^* - \omega' +i\epsilon)}, \\
X^A_2 &=& \frac{1}{(M_{f_1} - \omega^* - \omega +
i{\Gamma_{K^{*+}}}/{2})(E_V
- \omega - \omega' + i\epsilon)}, \\
X^A_3 &=& \frac{1}{(\omega + \omega^* - E_\gamma -
i{\Gamma_{K^{*+}}}/{2})(E_V + \omega + \omega'
 - i\epsilon)},
\end{eqnarray}
with $\omega = \sqrt{|\vec{q}~|^2 + m^2_{K^-}}$ and $\omega' =
\sqrt{|\vec{q}~|^2 + E^2_\gamma +2E_\gamma |\vec{q}~|cos\theta
+m^2_{K^+}}$ the energies of $K^-$ and $K^+$ in the diagram of
Fig.~\ref{Fig:feydiagram} $A)$. $F^B_1$ and $F^B_2$ will be obtained
just applying the substitution to $F^A_1$ and $F^A_2$ with
$m_{K^{*+}} \to m_{K^{*0}}$, $m_{K^-} \to m_{\bar{K}^0}$, and
$m_{K^+} \to m_{K^0}$.
The partial decay width takes the form
\begin{eqnarray}
\Gamma_{f_1(1285) \to \gamma V} &=&
\frac{e^2g^2g^2_{f_1}E^5_\gamma}{192\pi^2M^2_{f_1}m^2_V}
\sum_{i=1,2} |\int ^\Lambda_0 d|\vec{q}~| \int ^{1}_{-1}dcos\theta
\nonumber \\
&& \big(C_A F^A_i + C_B F^B_i \big )|^2, \label{eq:pdw}
\end{eqnarray}
with
\begin{eqnarray}
C_A &=& -\frac{\sqrt{2}}{4} \frac{g_{K^{*+} K^+
\gamma}}{m_{K^{*+}}}, ~~ {\rm for} ~ V=  \rho^0 ,  \omega, \\
C_A &=& \frac{1}{2} \frac{g_{K^{*+} K^+
\gamma}}{m_{K^{*+}}}, ~~ {\rm for} ~ V=  \phi, \\
C_B &=& \frac{\sqrt{2}}{4} \frac{g_{K^{*0} K^0
\gamma}}{m_{K^{*0}}}, ~~ {\rm for} ~ V=  \rho^0, \\
C_B &=& -\frac{\sqrt{2}}{4} \frac{g_{K^{*0} K^0
\gamma}}{m_{K^{*0}}}, ~~ {\rm for} ~ V=   \omega, \\
C_B &=& -\frac{1}{2} \frac{g_{K^{*0} K^0 \gamma}}{m_{K^{*0}}}, ~~
{\rm for} ~ V=  \phi.
\end{eqnarray}

For $\rho^0$ production, the relative minus sign between $C_A$ and $C_B$ combined with the minus sign between the couplings $g_{K^{*+} K^+
\gamma}$ and $g_{K^{*0} K^0
\gamma}$ is positive, and hence the interference of the two diagrams $A)$ and $B)$ shown in Fig.~\ref{Fig:feydiagram} is constructive. However, it is  destructive for $\omega$ and $\phi$ production, which will make the $\Gamma_{f_1(1285) \to \gamma \rho^0}$ is much lager than the other two partial decay widths.

\section{Numerical results and discussion}  \label{sec:results}

\begin{figure}[htbp]
\centering
\includegraphics[scale=0.5]{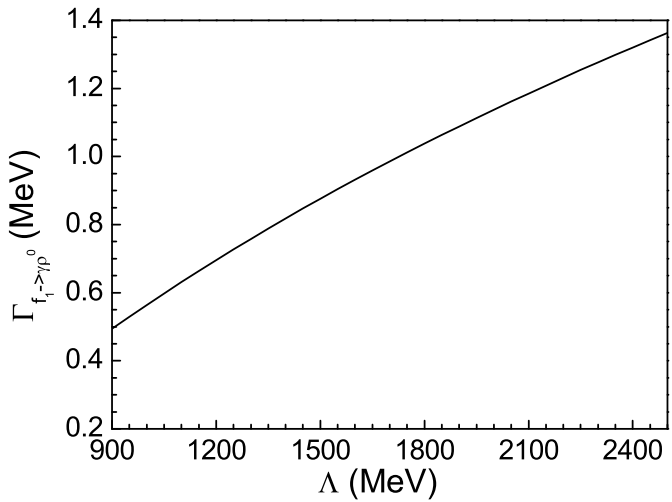}
\caption{Partial decay width of $f_1(1285) \to \gamma \rho^0$ decay
as a function of the cutoff parameter $\Lambda$.}
\label{Fig:pdwrho}
\end{figure}

A momentum cutoff $\Lambda$ is introduced in Eq.~\eqref{eq:pdw}, and the partial decay
width of $f_1(1285) \to \gamma \rho^0$ decay as a function of the $\Lambda$ from $900$ to $2500$ MeV is illustrated in Fig.~\ref{Fig:pdwrho}. We can see that, in the range of cutoff we consider, the $\Gamma_{f_1(1285) \to \gamma \rho^0}$ varies from $0.5$ to $1.4$ MeV, which is consistent with the experimental result within errors~\cite{Dickson:2016gwc,Tanabashi:2018oca}. In table~\ref{tab:results}
we show explicitly the numerical results of the $f_1(1285) \to \gamma V$ decays with some particular cutoff parameters.

\begin{table}[htbp]
\caption{Partial decay width for $f_1(1285) \to \gamma V$. All units are in MeV.}\label{tab:results}
\begin{tabular}{|c|c|c|c|}
\hline
$\Lambda$   & $f_1 \to \gamma \rho^0$ & $f_1 \to \gamma \omega$ [$\times 10^{-2}$]  & $f_1 \to \gamma \phi$ [$\times 10^{-2}$] \\
\hline
 $1000$      & $0.56$  & $1.87$  & $0.93$    \\
 $1500$      & $0.88$  & $3.01$   & $1.40$    \\
 $2000$      & $1.14$   & $4.01$  & $1.78$    \\
 $2500$     & $1.36$   & $4.87$  & $2.09$    \\
 Exp.~\cite{Tanabashi:2018oca} & $1.18\pm 0.27$ & ---  & $1.70 \pm 0.61$ \\
\hline
\end{tabular}
\end{table}

In general we cannot provide the value of the cutoff parameter,
however, if we divide $\Gamma_{f_1(1285) \to \gamma \rho^0}$ by
$\Gamma_{f_1(1285) \to \gamma \omega}$ or $\Gamma_{f_1(1285) \to \gamma \phi}$,
the dependence of these ratios on the cutoff will be smoothed. Two ratios are defined as
\begin{eqnarray}
R_1 &=& \frac{\Gamma_{f_1(1285) \to \gamma \rho^0}}{\Gamma_{f_1(1285) \to \gamma
\phi}}, \nonumber \\
R_2 &=& \frac{\Gamma_{f_1(1285) \to \gamma \rho^0}}{\Gamma_{f_1(1285) \to \gamma
\omega}} . \label{eq:ratios}
\end{eqnarray}
These two ratios are correlated with each other. With $R_1$ measured by the experiment, one can fix the cutoff in the model and predict the ratio $R_2$.


\begin{figure}[htbp]
\centering
\includegraphics[scale=0.5]{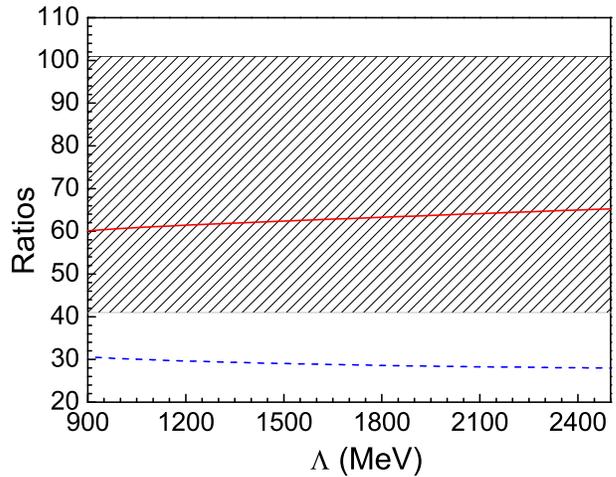}
\caption{The $\Lambda$ dependence of the ratios $R_1$ (solid line) and $R_2$ (dashed line) defined in Eq.~(\ref{eq:ratios}). The error band correspond the experimental result for $R_1$.} \label{Fig:ratio}
\end{figure}

In Fig.~\ref{Fig:ratio}, we show the numerical results for the
above ratios, where the solid line stands for the results for $R_1$,
while the dashed line stands for the results for $R_2$. Indeed, one sees that the dependence of both ratios on the cutoff is rather weak. The ratio $R_1 \simeq 60$ is in agreement with the experimental result $71 \pm 30$~\cite{Tanabashi:2018oca}. On the other hand, the result of $R_2$ is about $30$. It is a firm conclusion that the partial decay width of $f_1(1285) \to \gamma \rho^0$ is much larger than the ones to $\gamma \omega$ and $\gamma \phi$ channels. This is because the destructive interference between Fig.~\ref{Fig:feydiagram} $A)$ and $B)$ for $\omega$ and $\phi$ production. Our conclusion here is different with these quark model calculations~\cite{Osipov:2017ray,Ishida:1988uw}. We hope that the future experimental measurements can clarify this issue. Further theoretical
research considering both the molecular and $q\bar{q}$ components
for the $f_1(1285)$ state would be most welcome after the
discussion made here.

\section{Summary}

In this work, we evaluate the partial decay width of the radiative decays  $f_1(1285)
\to \gamma V$ with the assumption that the $f_1(1285)$ is
dynamically generated from the $\bar K^* K$ interaction. The results we obtained for the partial widths are compatible with
experimental data within errors. Furthermore, we find some relevant features of our model calculations, which turn out to be very
different from other theoretical predictions using quark models. The precise experimental observations of
those radiative decays would then provide very valuable
information on the relevance of components in the $f_1(1285)$
wave function.

\section*{Acknowledgments}

This work is partly supported by the
National Natural Science Foundation of China under Grant Nos. 11735003, 11675091, 11835015, and 11475227 and the Youth Innovation Promotion Association CAS (2016367).

\bibliographystyle{plain}

\end{document}